\documentclass[aps, superscriptaddress,nofootinbib, reprint]{revtex4-2}
\usepackage{graphicx}
\usepackage{xcolor}
\usepackage[%
colorlinks=true,
linkcolor=blue,
citecolor=blue,
]{hyperref}
\usepackage{booktabs}
\usepackage{rotating}
\usepackage[marginal]{footmisc}
\usepackage{graphicx}
\usepackage{dcolumn}
\usepackage{bm}
\usepackage{slashed} 
\usepackage{amssymb}
\newcommand{\Tr}{\operatorname{Tr}}

\usepackage{amsmath}
\allowdisplaybreaks[4] 

\usepackage[T1]{fontenc} 
\usepackage{tensor}
\usepackage{braket}  
\usepackage{bm}      
\usepackage{upgreek}
\usepackage{extarrows}

\newcommand\MaE{\mspace{2mu}\mathrm{e}\mspace{2mu}} 
\newcommand\MaPI{\mspace{2mu}\uppi\mspace{2mu}} 
\newcommand\MaD{\,\mathrm{d}} 

\begin{document}
\title{Axion Inflation from Heavy-Fermion One-Loop Effects}

\author{Kai-Ge Zhang}
\email{zhangkaige21@mails.ucas.ac.cn}
\affiliation{International Centre for Theoretical Physics Asia-Pacific,
University of Chinese Academy of Sciences, 100190 Beijing, China}
\affiliation{Taiji Laboratory for Gravitational Wave Universe, University of
Chinese Academy of Sciences, 100049 Beijing, China}

\author{Jian-Feng He}
\email{hejianfeng@itp.ac.cn}
\affiliation{Institute of
Theoretical Physics, Chinese Academy of Sciences (CAS), Beijing 100190, China}
\affiliation{School of Physical Sciences, University of Chinese Academy of
Sciences, No.19A Yuquan Road, Beijing 100049, China}

\author{Chengjie Fu}
\email{fucj@ahnu.edu.cn}
\affiliation{Department of Physics, Anhui Normal University, Wuhu, Anhui 241002, China}

\author{Zong-Kuan Guo}
\email{guozk@itp.ac.cn}
\affiliation{Institute of
Theoretical Physics, Chinese Academy of Sciences (CAS), Beijing 100190, China}
\affiliation{School of Physical Sciences, University of Chinese Academy of
Sciences, No.19A Yuquan Road, Beijing 100049, China}
\affiliation{School of Fundamental Physics and Mathematical Sciences,
Hangzhou Institute for Advanced Study, University of Chinese Academy of
Sciences, Hangzhou 310024, China}


\begin{abstract}
We derive a one-loop effective description of axion inflation by integrating out a heavy Dirac fermion with an inflaton-dependent complex mass undergoing a smooth localized threshold transition. The threshold induces correlated corrections to the inflaton and gauge sectors, including a Coleman-Weinberg term, a vacuum-polarization correction, and an anomaly-induced Chern-Simons coupling. Together, these effects transiently enhance and localize gauge-field production, generating a chiral stochastic gravitational-wave background in the deci-hertz band within the projected sensitivities of BBO and DECIGO, while remaining below representative primordial-black-hole bounds.
\end{abstract}
\maketitle

\section{Introduction}
\label{sec: introduce}
Inflation provides a compelling description of the very early Universe \cite{Guth:1980zm, Sato:1980yn, Linde:1981mu, Albrecht:1982wi, Starobinsky:1980te}, offering a unified explanation for the observed homogeneity, isotropy, and near scale invariance of primordial perturbations \cite{Starobinsky:1979ty, Mukhanov:1981xt, Hawking:1982cz, Guth:1982ec, Starobinsky:1982ee, Abbott:1984fp}, while resolving the horizon and flatness problems of the standard cosmological model~\cite{Brout:1977ix,Kazanas:1980tx}. In the simplest slow-roll realization~\cite{Linde:1981mu,Albrecht:1982wi}, an inflaton field drives accelerated expansion and its quantum fluctuations seed the CMB anisotropies. Achieving a successful inflationary history with $\mathcal{O}(60)$ e-folds~\cite{Planck:2018jri,Abbott:1982hn}, however, typically demands an exceptionally flat potential. In effective field theory this flatness is generically obstructed by radiative corrections~\cite{Coleman:1973jx}, which induce large contributions to the inflaton mass and spoil slow-roll dynamics~\cite{Stewart:1994ts,Dine:1995uk,Baumann:2007np}.

Axion-like fields offer a natural avenue to overcome these challenges.
As pseudo–Nambu–Goldstone bosons, axions enjoy an approximate shift symmetry that protects the flatness of their potential against quantum corrections~\cite{Freese:1990rb,Kim:2004rp,Dimopoulos:2005ac,Easther:2005zr,McAllister:2008hb,Flauger:2009ab,Anber:2009ua,Kaloper:2008fb}, making them theoretically well-motivated candidates for the inflaton.

A generic feature of axions is their coupling to gauge fields through
Chern--Simons interactions~\cite{Namba:2015gja, Campeti:2022acx, Barnaby:2012xt, Ozsoy:2017blg, Garcia-Bellido:2016dkw}. As demonstrated in a series of works
\cite{Barnaby:2010vf, Barnaby:2011vw, Meerburg:2012id, Sorbo:2011rz,
Cook:2011hg, Barnaby:2011qe, Linde:2012bt, Dimopoulos:2012av,
Urban:2013spa}, the rolling axion induces a tachyonic instability in one
helicity of the gauge field, characterized by the dimensionless
parameter $\xi$. This instability leads to an exponential amplification
of gauge-field fluctuations, scaling as $\propto e^{\pi \left| \xi \right|}$. The amplified gauge modes can leave a broad range of observational signatures, including parity violation in the correlators~\cite{Sorbo:2011rz,Shiraishi:2013kxa,Philcox:2023xxk,Fujita:2023inz,Stefanyszyn:2023qov,Niu:2022fki,Ozsoy:2021onx}, CMB non-Gaussianity~\cite{Durrer:2024ibi,Philcox:2023xxk,Barnaby:2010vf, Barnaby:2011vw,Ozsoy:2021onx, Barnaby:2011qe,Jamieson:2025ngu,Dimastrogiovanni:2023juq,Linde:2012bt}, scale-dependent enhancements of the scalar power spectrum (potentially leading to the formation of primordial black holes, PBHs~\cite{Linde:2012bt,Talebian:2022jkb,Ozsoy:2023ryl,Talebian:2022cwk,Garcia-Bellido:2016dkw}), and chiral gravitational waves on both CMB and interferometer scales~\cite{Ferreira:2014zia, Namba:2015gja,Garcia-Bellido:2023ser, Garcia-Bellido:2016dkw,Zhang:2025cyd,Corba:2025reo,Corba:2024tfz,Lozanov:2023rcd,Unal:2023srk,Niu:2023bsr,barbon2025axion,Niu:2022quw}.
A number of extensions have also been explored. These include (i) setups in which the axion is not the inflaton but instead plays the role of a spectator~\cite{Namba:2015gja, Campeti:2022acx, Ozsoy:2017blg, Mukohyama:2014gba, Barnaby:2012xt} or auxiliary field~\cite{Zhang:2025dwp}; (ii) modifications of the axion sector itself, such as deformations of the potential ~\cite{Garcia-Bellido:2016dkw,Lizarraga:2025aiw,Talebian:2025jeg} (e.g., axion monodromy~\cite{Ozsoy:2020ccy,Ozsoy:2020kat,Cheng:2018yyr}), noncanonical kinetic terms~\cite{Michelotti:2024bbc,Kume:2025lvz}, or nonminimal axion–gravity couplings~\cite{Domcke:2017fix}; and (iii) the inclusion of additional degrees of freedom, most notably Schwinger-produced charged particles~\cite{iarygina2025schwinger,Domcke:2018eki,vonEckardstein:2024tix,Fujita:2022fwc}. 
More recently, extensions of the gauge sector itself have also been considered, including nonminimal gauge–gravity couplings~\cite{He:2024bno} and field-dependent gauge kinetic and Chern--Simons terms \cite{Durrer:2024ibi,Demozzi:2009fu,Fujita:2013qxa,Bamba:2003av,Bamba:2006ga,teuscher2026gravitational,Barnaby:2011qe}. 

Although phenomenologically rich, most of the extensions outlined above are formulated in a bottom-up spirit and driven primarily by phenomenological viability. In many cases~\cite{Domcke:2017fix,Garcia-Bellido:2016dkw,He:2024bno,Zhang:2025dwp}, not only the relevant couplings and field content, but also the axion-potential features invoked to trigger a transient fast-roll phase, are introduced directly at the effective level and treated as largely independent ingredients. This raises a basic question: can such correlated structures instead emerge from a UV-motivated construction, rather than being imposed by hand?
This question is particularly well motivated because the axion–gauge Chern--Simons interaction is not a generic infrared embellishment~\cite{Reece:2023iqn}, but a topological term whose coefficient is fixed by UV anomaly matching~\cite{Fujikawa:1979ay,tHooft:1980xss,Agrawal:2022lsp}.
As a first step toward a top-down understanding of axion–gauge dynamics, we consider a heavy charged Dirac fermion with an inflaton-dependent complex mass. Integrating out the fermion at one loop simultaneously generates an anomaly-induced Chern--Simons term and parity-even threshold corrections, leading to correlated structures in the effective inflaton potential and the gauge-sector couplings.

This paper is organized as follows. In Sec.~\ref{sec: MODEL}, we summarize the low-energy effective action used in the phenomenological analysis. In Secs.~\ref{sec: Background Dynamics}--\ref{sec: GRAVITATIONAL WAVES}, we study the background evolution, gauge-field production, and the resulting sourced perturbations within this EFT. Finally, in Sec.~\ref{sec: conclusion}, we present our conclusions. The technical details of the one-loop derivation from the UV fermion theory are deferred to Appendix~\ref{app:oneloop}.

Throughout this paper, we adopt natural units with $\hbar = c = 1$ and define the reduced Planck mass as $M_{\mathrm{pl}}\equiv (8 \MaPI G)^{-1/2}$. The symbol $t$ denotes the cosmic time, while $\tau$ denotes the conformal time. A dot, as in $\dot{\phi} \equiv \mathrm{d} \phi / \mathrm{d} t$, indicates a derivative with respect to the cosmic time, and a prime, as in $\phi' \equiv \mathrm{d} \phi / \mathrm{d} \tau$ indicates a derivative with respect to the conformal time. The scale factor is $a(t)$ and the Hubble parameter is $H \equiv \dot{a} / a$. Our sign convention of metric is $(- + + +)$.

\section{MODEL}
\label{sec: MODEL}
The one-loop derivation from the heavy-fermion UV completion is presented in Appendix~\ref{app:oneloop}. Retaining only the leading operators relevant for the dynamics studied below, we consider the resulting low-energy effective action
\begin{align}
S_{\rm eff}
=& \int \MaD^4x\,\sqrt{-g}\biggl[
\frac{M_{\mathrm{pl}}^2}{2}R
-\frac{1}{2}\partial^\mu\phi\,\partial_\mu\phi
-\tilde V(\phi)\nonumber\\
&\hspace{3.2em}
-\frac{1}{4}I_1(\phi)\,F^{\mu\nu}F_{\mu\nu}
-\frac{1}{4}I_2(\phi)\,F_{\mu\nu}\tilde F^{\mu\nu}
\biggr],
\label{eq:L_axion}
\end{align}
where $\phi$ is the axion inflaton and $F_{\mu\nu}\equiv \partial_\mu A_\nu-\partial_\nu A_\mu$ is the field strength of the U(1) gauge field. The dual field strength is defined as $\tilde{F}^{\mu\nu} \equiv \frac{1}{2}\eta^{\mu\nu\alpha\beta}F_{\alpha\beta}/\sqrt{-g}$, with the totally antisymmetric tensor $\eta^{\mu\nu\alpha\beta}$ normalized by $\eta^{0123} = 1$.

The EFT functions in Eq.~\eqref{eq:L_axion} arise from two distinct one-loop effects of the heavy fermion. The parity-even threshold corrections generate both a Coleman--Weinberg deformation of the underlying inflaton potential and a vacuum-polarization correction to the gauge kinetic term, which are encoded in $\tilde V(\phi)$ and $I_1(\phi)$, respectively. The anomaly-induced parity-odd contribution, in turn, generates the coefficient function $I_2(\phi)$ of the Chern--Simons term. Accordingly, we write
\begin{equation}
I_1(\phi)=1+\delta\,\left(T_S(\phi)-T_S(\phi_\ast)\right),
\label{eq:I1}
\end{equation}
\begin{equation}
I_2(\phi)=\alpha\,\left(T_P(\phi)-T_P(\phi_\ast)\right),
\label{eq:I2}
\end{equation}
\begin{equation}
\tilde V(\phi)=V(\phi)
+\gamma \,\left(T_S(\phi)-T_S(\phi_\ast)\right),
\label{eq:Veff}
\end{equation}
where $T_S(\phi)$ and $T_P(\phi)$ are two independent dimensionless profile functions of the heavy-fermion mass, introduced in Eqs.~\eqref{eq:mass1}--\eqref{eq:mass2}. Here $\phi_\ast$ is an arbitrary matching point, while $\delta$, $\alpha$, and $\gamma$ set the amplitudes of the induced contributions to the gauge kinetic term, the Chern--Simons coupling, and the deformation of the underlying inflaton potential, respectively. In this form, the EFT makes explicit that the feature in $\tilde V(\phi)$ and the field dependence of the gauge-sector couplings originate from the same heavy threshold, without assuming specific forms for $T_S(\phi)$ and $T_P(\phi)$.

\begin{figure}[tbp]
  \includegraphics[width=.45\textwidth]{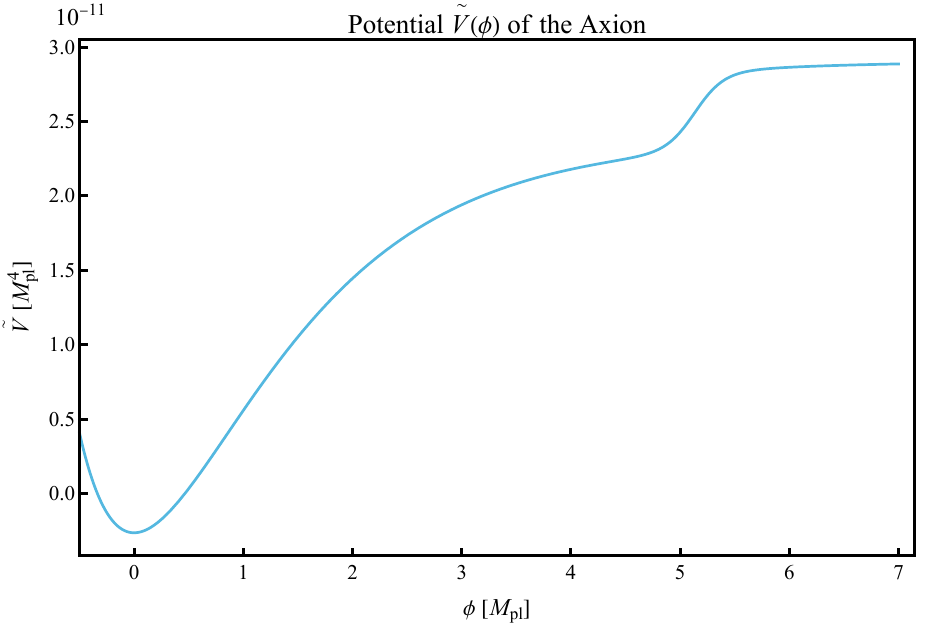}
  \caption{
Illustrative shape of the effective inflaton potential $\tilde V(\phi)$.
The potential exhibits a Starobinsky-like plateau at large field values,
while the threshold-induced transition occurs around $\phi\sim\phi_p$.
Such transient deformations modify the inflaton velocity only over a limited
range of e-folds and do not spoil the overall slow-roll evolution.
}
  \label{fig:chi_U}
\end{figure}

For phenomenological concreteness, we adopt the aligned smooth-step profile
\begin{equation}
T_S(\phi)=T_P(\phi)=\tanh\!\big[\beta(\phi-\phi_p)\big].
\label{eq:tanh-profile}
\end{equation}
This choice provides a minimal realization of a localized threshold crossing in the heavy-fermion sector and is also motivated by kink-like backgrounds, where tanh-type mass deformations arise naturally in fermionic systems with varying scalar backgrounds~\cite{Charmchi:2014kwa,Prokopec:2013ax}. We further choose $\phi_\ast=\phi_p$, so that the threshold corrections vanish at the center of the transition. To complete the specification of the model, we take the underlying inflaton potential to be the Starobinsky form~\cite{Starobinsky:1980te,Barrow:1988xh},
\begin{equation}
V(\phi)=V_0\!\left(1-e^{-\sqrt{2/3}\,\phi}\right)^{\!2},
\label{eq:V-flat}
\end{equation}
and choose $V_{0}=2.64\times10^{-11}M_{\mathrm{pl}}^4$, $\gamma=2.64\times10^{-12}M_{\mathrm{pl}}^4$, $\delta=0.9$, $\alpha=8.8$, $\beta=3.75M_{\mathrm{pl}}^{-1}$, and $\phi_{p}=5.135M_{\mathrm{pl}}$, whose relation to the underlying heavy-fermion parameters is summarized in Appendix~\ref{app:benchmark}. The resulting effective potential $\tilde V(\phi)$ is shown in Fig.~\ref{fig:chi_U}.

\section{Background Dynamics}
\label{sec: Background Dynamics}

From the action given in Eq.~\eqref{eq:L_axion}, along with the spatially flat Friedmann-Lemaître-Robertson-Walker metric, $\mathrm{d}s^{2} = -\mathrm{d}t^{2} + a^{2}(t) \mathrm{d}\bm{x}^{2}$, we can derive the following background equations,
\begin{align}
  \label{eq: a_eom_1}
  & H^{2} = \frac{1}{3 M_{\mathrm{pl}}^2} \rho, \\
  \label{eq: a_eom_2}
  & \frac{\ddot{a}}{a} + \frac{1}{2} \left( \frac{\dot{a}}{a}
  \right)^{2}
  = - \frac{1}{2 M_{\mathrm{pl}}^2} P, \\
  \label{eq: phi_eom}
  & \ddot{\phi} + 3 H \dot{\phi} + \tilde V_{,\phi}  = \frac{1}{2}I_{1,\phi}\braket{\bm{E}^{2} - \bm{B}^{2}}+I_{2,\phi}\braket{\bm{E} \cdot \bm{B}}, 
\end{align}
where the total energy density $\rho$ and the pressure $P$ are given by
\begin{align}
\label{eq: rho_tot}
  \rho =& \frac{1}{2} \dot{\phi}^{2}
  + \tilde V(\phi)  + \frac{I_{1}(\phi)}{2}\braket{\bm{E}^{2} + \bm{B}^{2}} , \\
\label{eq: p_tot}
  P =& \frac{1}{2} \dot{\phi}^{2} 
 - \tilde V(\phi) + \frac{I_{1}(\phi)}{6} \braket{\bm{E}^{2} + \bm{B}^{2}}.
\end{align}
Here, $\bm{E}$ and $\bm{B}$ are the electric and magnetic fields associated with the gauge field, defined by \cite{Barnaby:2010vf,Sorbo:2011rz}
\begin{align}
  \label{eq: E_def}
  & E_{i}(t) = - \dot{A}_{i} / a, \\
  \label{eq: B_def}
  & B_{i}(t) = \epsilon_{ijk} \partial_{j} A_{k} / a^{2}.
\end{align}
The angle bracket $\langle \cdots \rangle$ in Eqs.~\eqref{eq: phi_eom}-\eqref{eq: p_tot} denotes the ensemble average of the fields, which encodes backreaction of the quantum field on the classical background~\cite{Ballardini:2019rqh, Peloso:2022ovc, Domcke:2020zez, Byrnes:2011aa, Cheng:2015oqa, Notari:2016npn, DallAgata:2019yrr, Gorbar:2021rlt, Durrer:2023rhc,vonEckardstein:2023gwk, Iarygina:2023mtj, Galanti:2024jhw, vonEckardstein:2025oic, Anber:2009ua, Barnaby:2011qe, Sobol:2019xls, He:2024bno, He:2025ieo, Zhang:2025cyd, Caravano:2022epk, Caravano:2024xsb, Sharma:2024nfu, Figueroa:2024rkr, Lizarraga:2025aiw, Jamieson:2025ngu, Caravano:2021bfn, Caravano:2022epk, Figueroa:2023oxc}.

We decompose the gauge field $A_{i}(\tau, \bm{x})$ in the Fourier space as
\begin{align}
  A_{i}(\tau, \bm{x})
  = \int \frac{\mathrm{d}^{3}k}{(2\pi)^{{3}/{2}}}
  \sum_{\lambda=+,-} \varepsilon_{i}^{\lambda}(\hat{k}) A^{\lambda}(\tau, \bm{k})
  \MaE^{i \bm{k} \cdot \bm{x}},
\label{eq: Aq_Fourier}
\end{align}
where $\lambda = +, -$ denotes the polarization, $\hat{k} \equiv \bm{k}/k$ is a unit vector, $k \equiv |\bm{k}|$ is the norm, and $\epsilon_{i}^{\lambda}(\hat{k})$ is the circular polarization vector basis satisfying
\begin{align}
\label{eq: circular_1}
  & \bm{\varepsilon}^{\lambda}(\hat{{k}})
  = \left( \bm{\varepsilon}^{\lambda}(-\hat{{k}}) \right)^*, \\
  & \left( \bm{\varepsilon}^{\lambda}(\hat{{k}}) \right)^*
  \cdot \bm{\varepsilon}^{\lambda'}(\hat{{k}})
  = \delta_{\lambda \lambda'}, \\
  &\bm{{k}} \cdot \bm{\varepsilon}^{\lambda}(\hat{{k}}) = 0, \\
\label{eq: circular_4}
  &\bm{{k}} \times \bm{\varepsilon}^{\lambda}(\hat{{k}})
  = (-\lambda) \mathrm{i} k \, \bm{\varepsilon}^{\lambda}(\hat{{k}}).
\end{align}
We perform canonical quantization by promoting the classical field $A^{\lambda}(\tau,\bm{k})$ to the operator $\hat{A}^{\lambda}(\tau,\bm{k})$,
\begin{equation}
\label{eq: Aq_quantization}
\hat{A}^{\lambda}(\tau,\bm{k}) = A^{\lambda}(\tau,\bm{k})\hat{a}_{\lambda}(\bm{k}) + \left( A^{\lambda}(\tau,-\bm{k}) \right)^{\!*} \hat{a}_{\lambda}^{+}(-\bm{k}),
\end{equation}
where $\hat{a}_{\lambda}(\bm{k})$ and $\hat{a}_{\lambda}^{+}(\bm{k})$ are annihilation and creation operators, respectively, satisfying the standard commutation relations,
\begin{align}
  \label{eq: a_commut_1}
  \left[ \hat{a}_{\lambda}(\bm{k}), \hat{a}_{\lambda'}^{\dagger}(\bm{k}') \right]
  &= \delta_{\lambda\lambda'} \delta^{(3)} (\bm{k} - \bm{k}'), \\
  \label{eq: a_commut_2}
  \left[ \hat{a}_{\lambda}^{\dagger}(\bm{k}), \hat{a}_{\lambda'}^{\dagger}(\bm{k}') \right]
  &= 0, \\
  \label{eq: a_commut_3}
  \left[ \hat{a}_{\lambda}(\bm{k}), \hat{a}_{\lambda'}(\bm{k}') \right]
  &= 0.
\end{align}
Then, by the definition of $\bm{E}$ and $\bm{B}$, one can obtain
\begin{align}
  \label{eq: ensemble_EB}
  & \braket{\bm{E} \cdot \bm{B}} =
  - \frac{1}{4 \MaPI^{2} a^{4}}
  \sum_{\lambda=\pm} \lambda \int_{0}^{\infty}
  \mathrm{d}k k^3
  \frac{\mathrm{d} }{\mathrm{d} \tau} \left|A^{\lambda}(\tau,\bm{k})\right|^{2}, \\
  \label{eq: ensemble_EE}
  & \langle E^{2}\rangle=\frac{1}{2\pi^{2}a^{4}}\sum_{\lambda=\pm}\int_{0}^{\infty}\mathrm{d}k\, k^{2} \left| \frac{\mathrm{d}A^{\lambda}(\tau,\bm{k})}{\mathrm{d}\tau} \right|^{2}, \\
  \label{eq: ensemble_BB}
  & \braket{B^{2}} = \frac{1}{2 \MaPI^{2} a^{4}}
  \sum_{\lambda=\pm} \int_{0}^{\infty} \mathrm{d}k k^{4}
  |A^{\lambda}(\tau,\bm{k})|^{2}.
\end{align}
\section{PARTICLE PRODUCTION}
\label{sec: PARTICLES PRODUCTION}
Substituting the mode decomposition in Eqs.~\eqref{eq: Aq_Fourier} and~\eqref{eq: Aq_quantization} into the action \eqref{eq:L_axion} and varying with respect to the gauge field yields the mode equation for the two circular polarizations
~\cite{Sorbo:2011rz},
\begin{equation}
\label{eq: Aq}
\ddot{A}^{\pm}_{\bm{k}} + \left(H+\frac{{I}_{1,\phi}}{I_1} \dot{\phi}\right)\dot{A}^{\pm}_{\bm{k}}
+ \left(\frac{k^2}{a^2}\mp\frac{k}{a}\frac{I_{2,\phi}}{I_1} \dot{\phi} \right) A^{\pm}_{\bm{k}} = 0,
\end{equation}

To quantify particle production it is convenient to introduce the canonically normalized variable
\begin{equation}
v_\lambda(\tau,k) \equiv \sqrt{I_1}\, A_\lambda(\tau,k),
\end{equation}
then the mode equation becomes
 \begin{equation}
v_{\lambda}''(\tau, k) + \omega_{\lambda}^{2}(\tau, k) v_{\lambda}(\tau, k)=0
\end{equation}
with the time-dependent frequency
 \begin{equation}
 \label{eq: massive frequnency}
 \omega_{\lambda}^{2}(\tau, k)=k^{2}-\lambda k\frac{I_{2}^{\prime}}{I_{1}}-\frac{\left(\sqrt{I_{1}}\right)^{\prime\prime}}{\sqrt{I_{1}}}. 
\end{equation}
Deep inside the horizon we impose the Bunch--Davies initial condition,
\begin{equation}
\left. v_{\pm}(\tau, k) \right|_{- k \tau \gg 1}
= \frac{1}{\sqrt{2k}} \mathrm{e}^{- \mathrm{i} k \tau } .
\end{equation}

The last two terms in Eq.~\eqref{eq: massive frequnency} encode the breaking of conformal invariance; whenever $\omega_{\lambda}^{2}(\tau,k)$ becomes negative, the corresponding helicity mode is excited and particles are produced. The helicity-dependent term,
\begin{equation}
-\lambda k\frac{I_{2}^{\prime}}{I_{1}},
\end{equation}
originates from the Chern–Simons coupling $I_{2}(\phi)\, F \tilde{F}$ and acts as an effective “chemical potential” for the two different circular polarizations. 
Its strength is conveniently characterized by
\begin{equation}
\xi_{\rm eff}
\equiv \frac{1}{2H}\frac{I_{2,\phi}}{I_1}\dot{\phi},
\end{equation}
so that for $\xi_{\rm eff}\neq 0$ one helicity develops a tachyonic instability and is exponentially amplified, $A_\lambda\propto e^{\pi\left|\xi_{\rm eff}\right|}$, while the opposite helicity remains close to its vacuum configuration~\cite{Barnaby:2011vw,Ozsoy:2020ccy}.
By contrast, the helicity-blind term,
\begin{equation}
-\frac{\left(\sqrt{I_{1}}\right)^{\prime\prime}}{\sqrt{I_{1}}},
\end{equation}
is induced by the time dependence of the gauge kinetic term $I_{1}(\phi)FF$ and plays the role of a pump field~\cite{Bamba:2003av,Bamba:2006ga}. A sufficiently rapid variation of $I_{1}(\phi)$ can therefore lead to helicity-symmetric amplification of gauge fluctuations.

\begin{figure}[t]
  \centering
  \includegraphics[width=0.9\linewidth]{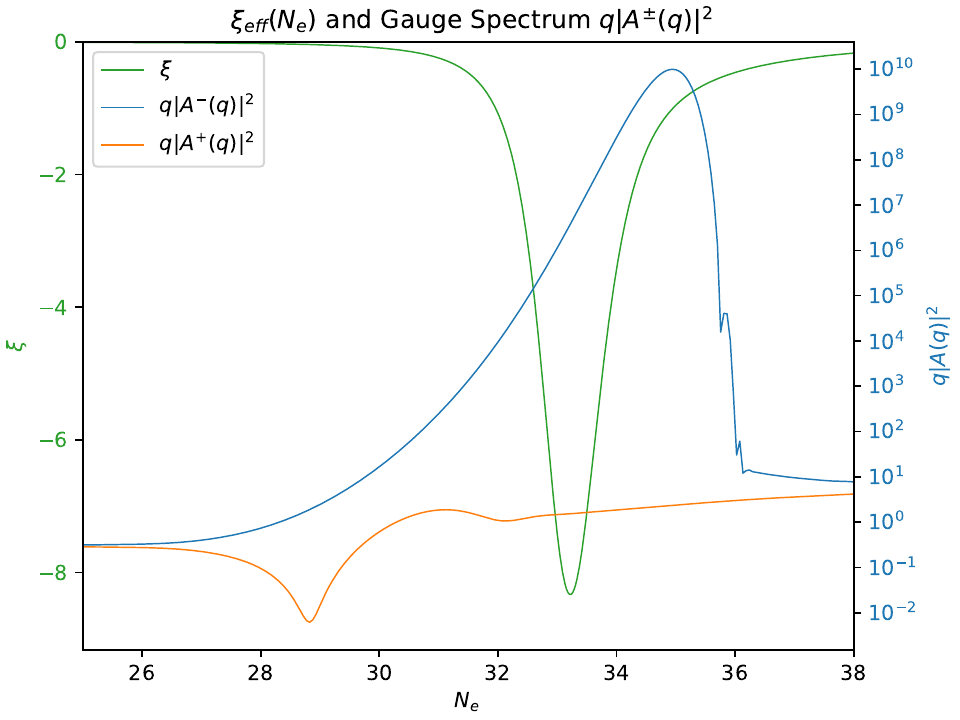}
  \caption{
Evolution of $\xi_{\text{eff}}(N_{e})$ and the gauge-field spectra $q|A^{\pm}(q)|^{2}$ for the benchmark threshold profile. The strong hierarchy $q|A^{-}|^{2}\gg q|A^{+}|^{2}$ identifies the transient, localized amplification as a helicity-dependent Chern--Simons instability, which shuts off once $I_{2,\phi}$ decays. The residual asymmetry across the feature is then controlled by the $I_{1}(\phi)$ normalization, $A_{\lambda}=v_{\lambda}/\sqrt{I_{1}}$, enhancing modes exiting for $I_{1}<1$ and suppressing earlier modes with $I_{1}>1$.
  }
  \label{fig:gauge_spectrum}
\end{figure}

For the benchmark smooth-step choice introduced in Sec.~\ref{sec: MODEL}, the relevant derivatives $I_1'$ and $I_2'$ are sharply localized around the transition $\phi\simeq\phi_p$, with width $\Delta\phi\sim \beta^{-1}$. Gauge-field production is therefore transient and tied to the feature, as shown in Fig.~\ref{fig:gauge_spectrum}. Since the Chern--Simons and pump terms act differently (helicity-selective versus helicity-symmetric), we identify the dominant mechanism by comparing the two helicity mode amplitudes. The same figure further shows that, in the parameter range of interest, the Chern--Simons term drives the amplification, while the pump contribution yields a negligible enhancement for both helicities.

Accordingly, the primary role of $I_1(\phi)$ is not to drive production by itself, but to enhance the effective instability parameter through the factor $1/I_1$: when $I_1(\phi)$ becomes temporarily small, $I_1(\phi)<1$, $\xi_{\rm eff}$ is boosted relative to the canonical case ($I_1=1$), where efficient amplification typically requires parametrically large $\alpha$~\cite{Barnaby:2011vw,Anber:2009ua,Caravano:2024xsb,Namba:2015gja,Unal:2023srk}. In addition, since the vacuum is defined for $v_\lambda$, 
the physical mode $A_\lambda=v_\lambda/\sqrt{I_1}$ inherits an additional normalization: modes exiting while $\phi>\phi_p$ (so that $I_1<1$) are enhanced by $I_1^{-1/2}$, whereas modes exiting earlier at $\phi<\phi_p$ with $I_1>1$ are comparatively suppressed.

In addition, the coefficient function $I_2(\phi)$ acts as a dynamical switch for the Chern--Simons interaction. Across the transition it features an approximately constant slope,
\begin{equation}
I_{2,\phi}\simeq {\alpha}{\beta},
\end{equation}
so that the interaction reduces to the standard axion--Chern--Simons form, while away from the feature it is exponentially suppressed,
\begin{equation}
I_{2,\phi}\propto \operatorname{sech}^2\!\big[\beta(\phi-\phi_p)\big],
\end{equation}
effectively shutting off the tachyonic source. Amplification is therefore confined to a finite interval in field space (and hence to a limited number of e-folds), avoiding excessive late-time gauge production that would otherwise risk PBH overproduction~\cite{Niu:2023bsr,Cheng:2018yyr,Garcia-Bellido:2016dkw,Linde:2012bt}.

\section{GRAVITATIONAL WAVES}
\label{sec: GRAVITATIONAL WAVES}
Building on the background dynamics and gauge-field production discussed above, we now turn to the sourced scalar and tensor perturbations. Our numerical strategy follows our previous studies~\cite{He:2024bno,He:2025ieo,Zhang:2025cyd,Zhang:2025dwp}: we simultaneously evolve Eqs.~\eqref{eq: a_eom_1}--~\eqref{eq: phi_eom} and the gauge-field mode equation~\eqref{eq: Aq}, evaluate the backreaction terms at each time step through the Fourier-space integrals in Eqs.~\eqref{eq: ensemble_EB}--~\eqref{eq: ensemble_BB}, and then compute the corresponding momentum and time integrals that determine the spectra.

Tensor perturbations around the flat FLRW background are described by
\begin{equation}
\mathrm{d}s^{2}=a^{2}(\tau)\left[-\mathrm{d}\tau^{2}
+\left(\delta_{ij}+h_{ij}\right)\mathrm{d}x^{i}\mathrm{d}x^{j}\right],
\end{equation}
where $h_{ij}$ is transverse and traceless, i.e., $h_{ii} = \partial_{j}h_{ij} = 0$.

We then decompose tensor perturbations $\hat{h}_{ij}(\tau, \bm{x})$ in Fourier space as
\begin{equation}
\label{eq: hq_Fourier}
\begin{aligned}
& \hat{h}_{ij}(\tau, \bm{x}) \\
&= \frac{2}{M_{\mathrm{pl}}}\frac{1}{a(\tau)}
  \int \frac{\mathrm{d}^3 \bm{k}}{(2\pi)^{{3}/{2}}}
  \sum_{\lambda=+,-}
  \varepsilon_{i}^{\lambda}(\hat{k}) \varepsilon_{j}^{\lambda}(\hat{k})
  \hat{h}_{\lambda}(\tau, \bm{k})
  e^{i \bm{k} \cdot \bm{x}} \\
&= \frac{2}{M_{\mathrm{pl}}}\frac{1}{a(\tau)}
  \int \frac{\mathrm{d}^3 \bm{k}}{(2\pi)^{{3}/{2}}}
  \sum_{\lambda=+,-}
  \varepsilon_{i}^{\lambda}(\hat{k}) \varepsilon_{j}^{\lambda}(\hat{k}) \\
&\times \bigg( h_{\lambda}^{\mathrm{v}}(\tau, \bm{k}) \hat{b}_{\lambda}(\bm{k})
  + \Big( h_{\lambda}^{\mathrm{v}}(\tau, -\bm{k}) \Big)^{\!*} \hat{b}_{\lambda}^{\dagger}(-\bm{k})
  + \hat{h}_{\lambda}^{\mathrm{s}}(\tau, \bm{k}) \bigg) e^{i \bm{k} \cdot \bm{x}},
\end{aligned}
\end{equation}
where $\hat{b}_{\lambda}(\bm{k})$ and $\hat{b}_{\lambda}^{\dag}(\bm{k})$ are the annihilation and creation operators, respectively, satisfying the standard commutation relations; $\epsilon_{i}^{\lambda}(\hat{k})$ is the circular polarization vector basis defined in Eqs.~\eqref{eq: circular_1}-\eqref{eq: circular_4}.

${h}_{\lambda}^{\mathrm{v}}{}(\tau, \bm{k})$ is a homogeneous solution corresponding to the vacuum GWs, while $\hat{h}_{\lambda}^{\mathrm{s}}(\tau, \bm{k})$ is the sourced contribution, which obeys
\begin{equation}
\label{eq: hq}
\begin{aligned}
  & h_{\lambda}^{\mathrm{s}''}(\tau, \bm{k})
  + \left( k^{2} - \frac{a''}{a} \right) {h}_{\lambda}^{\mathrm{s}}(\tau, \bm{k})= -\frac{a^{3}I_1(\phi)}{M_{\mathrm{pl}}} \left( \varepsilon_{i}^{\lambda}(\hat{k}) \varepsilon_{j}^{\lambda}(\hat{k}) \right)^{\!*}
   \\
  &\times \int \frac{\mathrm{d}^{3} p}{(2\pi)^{3/2}}\left( \hat{E}_{i}(\tau,\bm{p}) \hat{E}_{j}(\tau,\bm{k}-\bm{p})
      + \hat{B}_{i}(\tau,\bm{p}) \hat{B}_{j}(\tau,\bm{k}-\bm{p}) \right).
\end{aligned}
\end{equation}
Such equations can be solved by the Green's function method.
Due to the tachyonic amplification of a single gauge-field helicity,
the source term on the RHS of Eq.~\eqref{eq: hq} is strongly chiral.
Consequently, the sourced gravitational waves inherit this helicity
asymmetry, leading to a parity-violating tensor spectrum.

The power spectrum of GWs, $\mathcal{P}^{\lambda}_{h}(\tau, k)$, is defined by
\begin{equation}
  \braket{\hat{h}_{\lambda}(\tau, \bm{k})\hat{h}_{\lambda}(\tau, \bm{k'})}
  \equiv \dfrac{2 \MaPI^{2}}{k^{3}}\dfrac{a(\tau)^2}{4}\mathcal{P}^{\lambda}_{h}(\tau,k) \delta(\bm{k} + \bm{k}').
\end{equation}

By applying Wick's theorem, the tensor power spectrum is given by
\begin{widetext}
\begin{align}
\notag
  & \mathcal{P}_{h}^{\lambda}(\tau,k)
  = \frac{H^{2}}{\pi^{2} M_{\mathrm{pl}}^{2}}
  + \frac{k^{3}}{\pi^{4} M_{\mathrm{pl}}^{4}} \int_{0}^{\infty} q^{2} \mathrm{d} q  \int_{-1}^{1} \mathrm{d} u \sum_{\lambda_1,\lambda_2=\pm}
  \Biggr[
    \left| \epsilon^{\lambda}_{i}(\hat{k}) \epsilon^{\lambda_1}_{i}(-\hat{q}) \right|^{2}
    \left| \epsilon^{\lambda}_{j}(\hat{k}) \epsilon^{\lambda_2}_{j}\left(\frac{\vec{q}-\vec{k}}{|\vec{q}-\vec{k}|}\right) \right|^{2} \\
  &\times \biggl|
  \int_{-\infty}^{0} \mathrm{d} \tau' I_1(\phi(\tau'))\frac{G_{k}(\tau, \tau')}{a(\tau)a(\tau')}
  \biggl(
    A^{'}_{\lambda_1}(\tau',\bm{q}) A^{'}_{\lambda_2}(\tau', \bm{k}-\bm{q})+ \lambda_1\lambda_2\, q |\bm{k} - \bm{q}|
    A_{\lambda_1}(\tau',\bm{q})
    A_{\lambda_2}(\tau',\bm{k} -\bm{q}) \biggr) \biggr|^{2} \Biggr],
\end{align}
\end{widetext}
where $u \equiv \cos \theta$, and the angular dependence of the polarization-vector contractions is given by
\begin{equation}
  \left| \epsilon_{i}^{\lambda}(\hat{p})
    \epsilon_{i}^{\lambda'}(\hat{q}) \right|^{2}
  = \left( \frac{1 - \lambda \lambda' \hat{p} \cdot \hat{q}}{2} \right)^{2},
\end{equation}
which explicitly encode the chiral
structure of the sourced gravitational waves,
and the Green's function is 
\begin{align}
  \notag
  G_{k}(\tau, \tau') &= \dfrac{1}{k^{3} \tau\tau'}
  \bigl[
    (1 + k^{2} \tau \tau') \sin(k (\tau - \tau')) \\
  \label{eq: green_func}
    & - k (\tau - \tau') \cos(k (\tau - \tau'))
  \bigr] \Theta(\tau - \tau').
\end{align}
The present-day GW energy density spectrum is related to the tensor power spectrum at the end of inflation by~\cite{Caprini:2018mtu}
\begin{equation}
\label{eq: omega_GW}
  \Omega_{\mathrm{GW},0}h^2 = \frac{\Omega_{\mathrm{r}, 0} h^{2}}{24}
  ( \mathcal{P}^{+}_{h}(\tau_{\mathrm{end}},k) + \mathcal{P}^{-}_{h}(\tau_{\mathrm{end}},k) ),
\end{equation}
where $\Omega_{\mathrm{r}, 0} \simeq 4.15 \times 10^{-5} / h^{2}$ denotes the current radiation abundance.
\begin{figure}[tbp]
  \includegraphics[width=.45\textwidth]{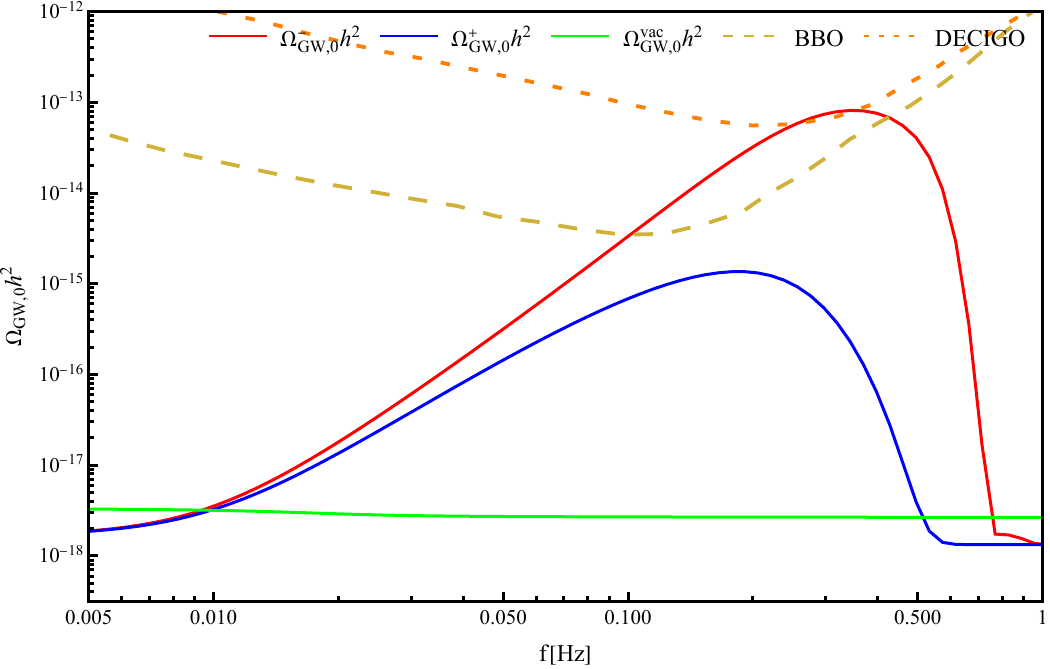}
  \caption{Present-day GW energy spectrum $\Omega_{\rm GW,0}^{\lambda}h^{2}$ for the benchmark model. The red, blue, and green curves show the $\lambda=-$, $\lambda=+$, and vacuum contributions, respectively, while the dashed lines indicate the projected sensitivities of BBO~\cite{Corbin:2005ny} and DECIGO~\cite{Seto:2001qf,Kawamura:2020pcg}. The transient threshold-induced instability enhances the chiral GW signal in the deci-hertz band, reaching $\Omega_{\mathrm{GW},0}h^2\sim 10^{-13}$, within the target sensitivity of future space-based interferometers.
 }
  \label{fig:GW}
\end{figure}

Gauge-field production also sources scalar perturbations.
Expanding the inflaton as $\phi=\phi_{0}+\delta\phi$ and retaining the
leading interaction terms, the canonically normalized variable
$Q_{\phi}\equiv a\,\delta\phi$ obeys
\begin{align}
&Q_{\phi}''+\left(k^{2}-\frac{a''}{a}\right)Q_{\phi}
= a^{3}\int \frac{\mathrm{d}^{3}\bm p}{(2\pi)^{3/2}}
\Bigg[
I_{2,\phi} E_i(\tau,\bm k-\bm p)B_i(\tau,\bm p)
\nonumber\\
&
+\frac{I_{1,\phi}}{2}
\Big(
E_i(\tau,\bm k-\bm p)\,E_i(\tau,\bm p)
-
B_i(\tau,\bm k-\bm p)\,B_i(\tau,\bm p)
\Big)
\Bigg].
\end{align}
where slow-roll suppressed metric contributions have been neglected.
The comoving curvature perturbation is given by
$\mathcal{R}\simeq -HQ_{\phi}/(a\dot{\phi})$ and
\begin{equation}
  \braket{\hat{\mathcal{R}}(\tau, \bm{k})\hat{\mathcal{R}}(\tau, \bm{k'})}
  \equiv \dfrac{2 \MaPI^{2}}{k^{3}}{P}_{\mathcal{R}}(\tau,k) \delta(\bm{k} + \bm{k}'),
\end{equation}
yielding the total scalar power spectrum
\begin{widetext}
\begin{align}
\mathcal{P}_{\mathcal{R}}(\tau,k)
&= \frac{H^{4}}{4\pi^{2}\dot{\phi}^{2}}
+ \frac{ k^{3} H^{2}}{16\pi^{4} \dot{\phi}^{2}}
\int_{0}^{\infty} q^{2}\,\mathrm{d}q
\int_{-1}^{1} \mathrm{d}u
\sum_{\lambda,\lambda'=\pm}
\Bigg\{
\left|
\epsilon_{i}^{\lambda}\!\left(\frac{\vec{k} - \vec{q}}{|\vec{k} - \vec{q}|}\right)
\epsilon_{i}^{\lambda'}(\hat{q})
\right|^{2}
\nonumber\\
&\quad \times
\left|
\begin{aligned}
&\int_{-\infty}^{\tau} \mathrm{d}\tau'\,
\frac{G_{k}(\tau,\tau')}{a(\tau)a(\tau')}
\Big[
\lambda'\,q\, A'_{\lambda}(\tau',\vec{k}-\vec{q})A_{\lambda'}(\tau',\vec{q})I_{2,\phi}(\phi(\tau')) 
+ \lambda\,|\vec{k}-\vec{q}|\,
A_{\lambda}(\tau',\vec{k}-\vec{q})A'_{\lambda'}(\tau',\vec{q})I_{2,\phi}(\phi(\tau'))\\
&-A'_{\lambda}(\tau',\vec{k}-\vec{q})A'_{\lambda'}(\tau',\vec{q})
I_{1,\phi}(\phi(\tau'))+\lambda'\,\lambda \,q\,|\vec{k}-\vec{q}|\,A_{\lambda}(\tau',\vec{k}-\vec{q})A_{\lambda'}(\tau',\vec{q})I_{1,\phi}(\phi(\tau'))\Big]
\end{aligned}
\right|^{2}
\Bigg\}.
\label{eq:P-R-1}
\end{align}
\end{widetext}

\begin{figure}[tbp]
  \includegraphics[width=.45\textwidth]{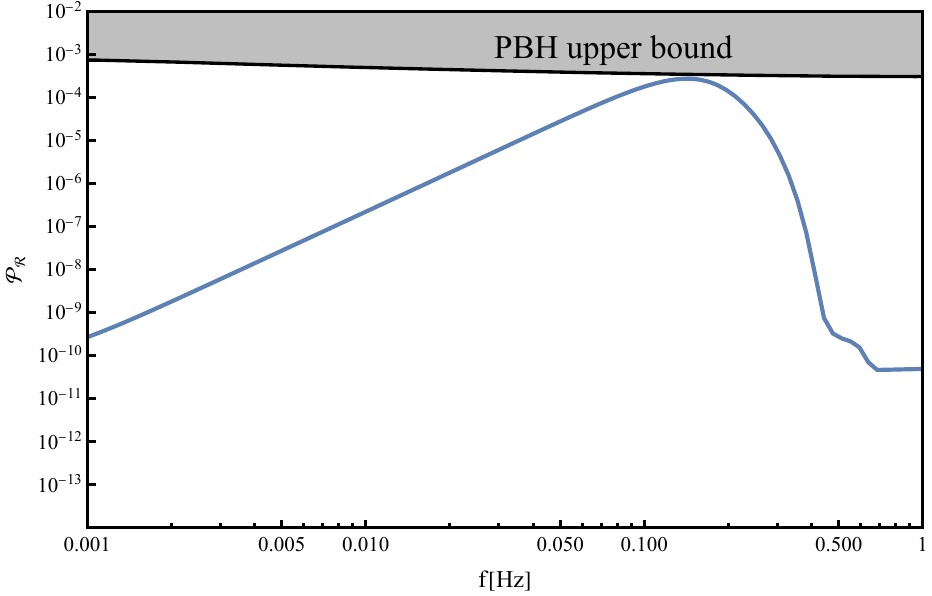}
  \caption{
 Curvature power spectrum ${\cal P}_{\cal R}$ sourced by the transient gauge-field instability. The gray shaded region denotes the increasingly stringent PBH upper bound~\cite{Lu:2019sti,Yi:2022ymw,Garcia-Bellido:2016dkw}. The spectrum peaks at deci-hertz frequencies while remaining below the PBH constraint.
 }
  \label{fig:Pr}
\end{figure}

The tensor and scalar spectra shown in Figs.~\ref{fig:GW} and~\ref{fig:Pr} are both sharply localized in frequency, indicating that the sourced signal is confined to a narrow band of scales.

This sharply contrasts with the standard axion--gauge setup~\cite{Cook:2011hg,Domcke:2016bkh,Barnaby:2011qe}, where a monotonic potential and an always-on Chern--Simons coupling typically strengthen the instability at later times, leading to excessive small-scale scalar power and PBH overproduction~\cite{Bugaev:2013fya,Erfani:2015rqv,Linde:2012bt,Ferreira:2014zia}. Although this tension may be partially alleviated by introducing a localized step in $V(\phi)$~\cite{Garcia-Bellido:2016dkw} to transiently enhance $\xi_{\text{eff}}$ at interferometer scales, such a feature is typically imposed by hand and does not prevent the renewed amplification of gauge fields as inflation approaches its end, $\epsilon_H\equiv-\dot H/H^2\to1$. The resulting late-time gauge-field production can drive the system into a strong backreaction regime and prolong the final stage of inflation by a few to $\mathcal{O}(10)$ e-folds~\cite{Barnaby:2011qe,Figueroa:2023oxc,Sharma:2024nfu}, thereby spoiling the mapping between the feature scale during inflation and the predicted interferometer-scale signal today. More importantly, if this late-time amplification persists into reheating, it may trigger gauge preheating and generate an additional chiral stochastic gravitational-wave background~\cite{Adshead:2018doq,Adshead:2019lbr}, which is constrained by current bounds on the extra radiation energy density~\cite{Planck:2018vyg,Smith:2006nka}.

In our framework, by contrast, the heavy-fermion threshold plays a dual role: the Coleman--Weinberg deformation encoded in $\tilde V(\phi)$ and the threshold correction to $I_{1}(\phi)$ jointly enhance gauge-field production around the feature, while $I_{2}(\phi)$ switches on the Chern--Simons interaction only in that range, thereby decoupling particle production at the target scales from that at other scales.

\section{Conclusions}
\label{sec: conclusion}

In this work, we have developed a UV-motivated effective description of axion--gauge inflation by integrating out a heavy Dirac fermion with an inflaton-dependent complex mass. At one loop, the heavy threshold generates a correlated EFT structure: parity-even corrections induce both a Coleman--Weinberg deformation of the inflaton potential $\tilde V(\phi)$ and a vacuum-polarization correction to the gauge kinetic function $I_1(\phi)$, while the parity-odd anomaly induces the Chern--Simons coupling $I_2(\phi)F\tilde F$. We have illustrated the resulting dynamics with a minimal smooth localized benchmark in which the threshold profiles are aligned in field space and share a common width.

The central physical consequence of this correlated threshold structure is that it reshapes the standard axion--gauge dynamics in a controlled and localized way. A transient suppression of $I_1(\phi)$ enhances $\xi_{\rm eff}$ and leads to efficient helicity-selective gauge amplification, while the localized variation of $I_2(\phi)$ acts as a dynamical switch that turns off the tachyonic source away from the threshold. Gauge production is therefore confined to a finite interval of e-folds, substantially reducing the usual late-time amplification problem. In addition, the threshold-induced deformation of $\tilde V(\phi)$ briefly accelerates the inflaton motion and further boosts $\xi_{\rm eff}$ near the interferometer scales of interest.

As a result, the model can generate a localized chiral stochastic gravitational-wave signal with peak amplitude $\Omega_{\mathrm{GW},0}h^2\sim10^{-13}$ in the deci-hertz band, within the projected reach of BBO and DECIGO, while remaining consistent with representative primordial-black-hole bounds. More broadly, our analysis shows that ingredients often introduced independently in phenomenological axion--gauge models can instead emerge in a correlated way from a common heavy threshold. It would be interesting to extend this framework to more explicit UV completions and to more general threshold profiles---including non-aligned, multiple, or asymmetric transitions---which may give rise to richer sourced scalar and tensor spectra, such as multi-peak or oscillatory structures.

\begin{acknowledgments}
This work is supported in part by the National Natural
Science Foundation of China under Grants No. 12475067,
No. 12305057, and No. 12235019.
\end{acknowledgments}

\appendix

\section{One-loop derivation of the effective action}
\label{app:oneloop}

In this appendix, we present the derivation of the effective theory used in Sec.~\ref{sec: MODEL} by integrating out the heavy charged Dirac fermion in the underlying UV completion.

We consider a UV theory in curved spacetime in which a heavy Dirac fermion $\Psi$ is coupled to an axion-like field $\phi$, which plays the role of the inflaton at low energies, and to a $U(1)$ gauge field $A_\mu$. The fermion-sector action takes the form
\begin{equation}
S_{\Psi}=\int d^4x \sqrt{-g}\;
\bar\Psi\left(i\slashed D - m_S(\phi)+ i\gamma_5 m_P(\phi)\right)\Psi,
\end{equation}
where \(m_S(\phi)\) and \(m_P(\phi)\) define the most general mass term without derivatives. Here $e^\mu_{\ a}$ is the inverse vierbein, satisfying $e^\mu_{\ a}e_{\mu}^{\ b}=\delta_a^{\ b}$ and $e^\mu_{\ a}e_{\nu}^{\ a}=\delta^\mu_{\ \nu}$, while the flat-space gamma matrices obey the Clifford algebra $\{\gamma^a,\gamma^b\}\equiv \gamma^a\gamma^b+\gamma^b\gamma^a=2\eta^{ab}$, with $\eta^{ab}=\mathrm{diag}(-,+,+,+)$, and we define $\gamma_5\equiv i\gamma^0\gamma^1\gamma^2\gamma^3$. The spinor covariant derivative is
\begin{equation}
D_\mu \equiv \partial_\mu + \frac{1}{8}\omega_{\mu ab}[\gamma^a,\gamma^b] + iqA_\mu,
\end{equation}
where $[\gamma^a,\gamma^b]\equiv \gamma^a\gamma^b-\gamma^b\gamma^a$, and $\omega_{\mu ab}=e_{a\nu}(\partial_\mu e_b^{\ \nu}+\Gamma^\nu_{\mu\rho}e_b^{\ \rho})$ is the spin connection associated with the vierbein, with $\Gamma^\nu_{\mu\rho}$ the Levi--Civita connection. Here $q$ denotes the $U(1)$ charge of the heavy fermion; more generally, it should be regarded as a model-dependent effective parameter of the heavy sector~\cite{Alonso-Alvarez:2018irt}. In simple UV completions, one typically 
expects $q={\cal O}(1)$, although values as large as ${\cal O}(10)$ may arise in multiplicity~\cite{Plakkot:2021xyx}, anomaly coefficients
~\cite{ Agrawal:2022lsp}, or large group-theoretic indices~\cite{DiLuzio:2020wdo}. The $\phi$-dependent mass functions induced by Yukawa couplings~\cite{Dorsch:2016nrg,Morrissey:2012db,Charmchi:2014kwa} are taken to be
\begin{align}
\label{eq:mass1}
m_S(\phi)&=M+\kappa\, T_S(\phi),\\
m_P(\phi)&=\sigma\, T_P(\phi).
\label{eq:mass2}
\end{align}
where the hard mass $M>0$ sets the overall heavy scale, while $T_S(\phi)$ and $T_P(\phi)$ are dimensionless profile functions encoding UV-dependent, shift-symmetry-breaking Yukawa structures and are left unspecified, with $\kappa$ and $\sigma$ setting their overall magnitudes. For an axion-like inflaton, approximate shift symmetry naturally favors the small-modulation regime $|\kappa|,|\sigma|\ll M$, so that the inflaton-dependent corrections to the fermion mass term $m(\phi)e^{-i\gamma_{5}\theta(\phi)}$ remain perturbative, with
\begin{equation}
m(\phi)\equiv\sqrt{m_S^2(\phi)+m_P^2(\phi)},\qquad\theta(\phi)=\arctan\!\frac{m_P(\phi)}{m_S(\phi)},
\end{equation}
This facilitates the realization of a heavy, adiabatic regime~\cite{Achucarro:2012sm,Kainulainen:2001cn,Kainulainen:2002th},  
\begin{equation}
m(\phi)\sim 10^3H,
\qquad
\left|\frac{m'(\phi)\,\dot\phi}{m^2}\right|\ll1,
\qquad
\left|\frac{\theta'(\phi)\,\dot\phi}{m}\right|\ll1,
\label{eq:heavy_adiabatic_regime}
\end{equation}
in which $\Psi$ can be integrated out consistently during inflation.

Then we can study the resulting one-loop corrections to the inflationary action
\begin{align}
  S_{\rm bos}=\int \!\! \MaD^4x\!  \sqrt{-g} \biggl[\!\
\frac{M_{\mathrm{pl}}^2}{2} R
- \frac{1}{2} \partial^\mu \phi \partial_\mu \phi- V(\phi)- \frac{1}{4} F^{\mu\nu} F_{\mu\nu}^{\phantom{\mu\nu}}
\!\biggr],
\label{eq:tree}
\end{align}
where $F_{\mu\nu}\equiv \partial_\mu A_\nu-\partial_\nu A_\mu$ is the $U(1)$ field strength. Here $V(\phi)$ denotes the underlying tree-level inflaton potential before integrating out the heavy fermion. At this stage it is left unspecified, except that it is assumed to be sufficiently flat, as expected for an axion-like inflaton with a weakly broken approximate shift symmetry.

In the path-integral formalism these corrections are encoded in the fermionic generating functional,
\begin{align}
Z_\Psi
\!=\!\int \!\mathcal{D}\Psi\mathcal{D}\bar{\Psi}
\exp\!\Biggl\{\!
\!-\!\int \!d^{4}x_{E}\sqrt{g_{E}}
\bar{\Psi}
\bigl(\slashed{D}_{E}\!+\!m(\phi)e^{-i\gamma_{5E}\theta(\phi)}\bigr)\Psi
\!\Biggr\},
\end{align}
where we have Wick-rotated to Euclidean signature (denoted by the subscript \(E\)). \footnote{Upon Wick rotating to Euclidean signature, \(x^0\to -i x_E^4\), we define \(A_4^{E}=iA_0\), \(A_i^{E}=A_i\), \(\gamma_E^4=i\gamma^0\), \(\gamma_E^i=\gamma^i\) \((i=1,2,3)\), and \(\gamma_{5E}\equiv -\gamma_E^1\gamma_E^2\gamma_E^3\gamma_E^4\).}
Then we remove the spacetime-dependent chiral phase of the fermion mass by a local axial rotation, $\Psi = e^{\frac{i}{2}\gamma_{5E}\theta(\phi)}\chi$, which renders the mass real and generates the familiar Fujikawa Jacobian $J_5[\theta;A_E,g_E]$ ~\cite{Fujikawa:1979ay,Fujikawa:1980eg,Endo:1984yn} responsible for the parity-odd terms (e.g., \(F\tilde F\)).
\begin{align}
Z_\Psi
\!=e^{i\,\Gamma_{\mathrm{odd}}}\int \mathcal{D}\chi \, \mathcal{D}\bar{\chi} \, 
\exp \Bigl\{-\int d^{4}x_{E} \, \sqrt{g_{E}} \, \bar{\chi}\,\mathcal{O}_{E}\, \chi \Bigr\} 
\end{align}
where
\begin{align}
\label{eq:odd-f}
\Gamma_{\mathrm{odd}}
&\equiv -i\,\ln J_{5}[\theta;A_E,g_E]\nonumber \\&= -\int d^{4}x_E \sqrt{g_{E}} \,\theta(\phi(x_E)) 
\lim_{s\to 0}\mathrm{tr}_V\!\left[
\gamma_{5E}\langle x_E|e^{-s\,\mathcal{H}_E}|x_E\rangle
\right]\\
\mathcal{H}_E& \equiv \mathcal{O}_E^\dagger \mathcal{O}_E \\
\mathcal{O}_{E}&\equiv \slashed{D}_{E}+m(\phi)+\frac{i}{2}\gamma_{E}^{\mu}\gamma_{5E}\partial_{\mu}^E\theta(\phi)
\end{align}
The second equality for $\Gamma_{\rm odd}$ is the standard Fujikawa representation~\cite{Fujikawa:1979ay,Adshead:2021ezw}, and $\mathrm{tr}_V$ denotes the trace over spinor (and gauge) indices at fixed $x_E$.

Since the $\chi$ path integral is Gaussian, it can be evaluated exactly, yielding a functional determinant up to an overall field-independent normalization of the measure,
\begin{equation}
\int \mathcal{D}\chi \,\mathcal{D}\bar{\chi}\;
\exp\!\left[-\!\int d^{4}x_{E}\sqrt{g_{E}}\;\bar{\chi}\,\mathcal{O}_{E}\,\chi\right]
= \det \mathcal{O}_{E}.
\end{equation}
To evaluate the determinant, we rewrite it as $
\det \mathcal{O}_E
=\exp\!\left[\frac{1}{2}\,\Tr\!\ln(\mathcal{H}_E)\right]
$, where $\Tr$ denotes the functional trace,
$\Tr(\cdots)\equiv\int d^4x_E\sqrt{g_E}\,\mathrm{tr}_V\langle x_E|(\cdots)|x_E\rangle$. We then define the parity-even part
\begin{align}
\Gamma_{\rm even}&\equiv -\,\frac{i}{2}\,\Tr\ln(\mathcal{H}_E)\nonumber\\
&= \frac{1}{2}\int_{0}^{\infty}\frac{ds}{s}\int d^{4}x_{E}\sqrt{g_{E}}\,
\mathrm{tr}_V\,\Big[\langle x_E|e^{-s\,\mathcal{H}_E}|x_E\rangle\Big]
\end{align}
We have used the proper-time identity $\Tr\ln \mathcal{H}_E = -\int_0^\infty \frac{ds}{s}\Tr(e^{-s\mathcal{H}_E})$.

Thus, 
\begin{equation}
Z_\Psi=\exp\!\left[\,i\,\Gamma_{\rm odd}+i\,\Gamma_{\rm even}\,\right].
\end{equation}
so that the inflationary effective action is
\begin{equation}
S_{\rm eff}=S_{\rm bos}+\Gamma_{\rm odd}+\Gamma_{\rm even}.
\end{equation}
To compute the one-loop radiative corrections, we employ the heat-kernel method~\cite{Vassilevich:2003xt,Karan:2017txu}, treating the parity-odd and parity-even contributions separately. For related flat-space analyses of one-loop QED-Yukawa effective actions in the constant-background approximation, see Ref.~\cite{Jacobson:2018kso}; here we work in curved spacetime and organize the result as a heavy-mass expansion for the local inflationary EFT.

\paragraph*{Parity-odd part (anomalous term).}
Performing the short-time (small-$s$) heat-kernel expansion of Eq.~\eqref{eq:odd-f} in four dimensions,
\begin{equation}
\lim_{s\to 0}\,\langle x_E|e^{-s\,\mathcal{H}_{E}}|x_E\rangle
=
\frac{1}{(4\pi s)^{2}}
\sum_{n=0}^{\infty} s^{n}\, a_{n}(x_E),
\end{equation}
where $a_n(x)$ are the Seeley--DeWitt coefficients~\cite{DeWitt:1964mxt,DeWitt:1967yk,DeWitt:1967ub,DeWitt:1967uc} \footnote{After rewriting the Euclidean operator in Laplace form, $\mathcal{H}_{E}=-(g_{E}^{\mu\nu}\nabla_{\mu}^{E}\nabla_{\nu}^{E}+\mathbb{E})$, with
$\nabla_{\mu}^{E}=D_{\mu}^{E}+\frac{i}{2}\gamma_{5E}\partial_{\mu}\theta(\phi)$
and $\mathbb{E}= -\frac{1}{4}R_{E}+\frac{i q}{4}[\gamma_{E}^{\mu},\gamma_{E}^{\nu}]F_{\mu\nu}^{E}-m^{2}(\phi)+\gamma_{E}^{\mu}\partial_{\mu}m(\phi)-i\,m(\phi)\gamma_{E}^{\mu}\gamma_{5E}\partial_{\mu}\theta(\phi)$, together with $\Omega_{\mu\nu}^E= i q F_{\mu\nu}^{E}-\frac{1}{4}\gamma_{E}^{\sigma}\gamma_{E}^{\rho}R_{\sigma\rho\mu\nu}^{E}$, the standard formulas of Ref.~\cite{Vassilevich:2003xt} Eqs.~(4.26)--(4.28) apply directly. Our $a_0$, $a_1$, and $a_2$ therefore correspond to Vassilevich's $a_0$, $a_2$, and $a_4$, respectively.},
we find that in four-dimensional spacetime the anomaly is governed by the $a_2$ term~\cite{Bastianelli:2018twb}. This yields
\begin{align}
\Gamma_{\rm odd}
\!=
\!\frac{-1}{16\pi^2}\!\int d^4x\sqrt{-g}\Bigg[q^{2}\theta(\phi)
F_{\mu\nu}\tilde F^{\mu\nu}\!+\!\frac{\theta(\phi)}{24}R_{\mu\nu\rho\sigma}\tilde R^{\mu\nu\rho\sigma}\Bigg].
\label{eq:odd}
\end{align}
Here the dual tensors are defined by $\tilde{F}^{\mu\nu} \equiv \eta^{\mu\nu\alpha\beta}F_{\alpha\beta}/\left(2\sqrt{-g}\right)$ and $\tilde{R}^{\mu\nu\rho\sigma}\equiv\eta^{\rho\sigma\alpha\beta}R^{\mu\nu}{}_{\alpha\beta}/\left(2\sqrt{-g}\right)$, where the totally antisymmetric tensor $\eta^{\mu\nu\alpha\beta}$ is normalized by $\eta^{0123} = 1$.

The first term in Eq.~\eqref{eq:odd} is the well-known Chern–Simons coupling \cite{Ellis:2020ivx,Quevillon:2021sfz,Adshead:2021ezw}. 
In the standard axion case~\cite{Barnaby:2010vf, Barnaby:2011vw, Barnaby:2011qe, Linde:2012bt, Planck:2015zfm, Ozsoy:2014sba}, an exact continuous shift symmetry fixes the phase to be linear, $\theta(\phi)\propto\phi$.
By contrast, heavy dynamics that break the shift symmetry can imprint nontrivial $\phi$-dependence through the fermion mass sector, leaving $\theta(\phi)$ model-dependent. In our setup,
\begin{equation}
\theta(\phi)\simeq\frac{\sigma}{M}T_{P}(\phi)+\mathcal{O}\bigg(\frac{\kappa\sigma}{M^{2}},\frac{\sigma^{3}}{M^{3}}\bigg).
\end{equation}

The second term in Eq.~\eqref{eq:odd} is the gravitational Chern--Pontryagin coupling~\cite{Jackiw:2003pm,Alexander:2009tp,Canizares:2012is}. Like the  Chern--Simons interaction, \(R_{\mu\nu\rho\sigma}\tilde R^{\mu\nu\rho\sigma}\) is a parity-odd pseudoscalar and thus vanishes on the homogeneous and isotropic FLRW background. At the perturbative level, however, the two couplings act very differently. In the gauge-field equation, the Chern--Simons interaction enters as a helicity-dependent tachyonic mass that competes with the redshifting quantity $ k^{2}/a^{2} $, and may therefore induce exponential growth. The gravitational Chern--Pontryagin term, by contrast, modifies only the coefficient of the friction term~\cite{Fu:2020tlw,Alexander:2004wk}, namely a constant $ \mathcal{O}(1) $ part of the tensor dynamics. In the controlled heavy, adiabatic regime~\eqref{eq:heavy_adiabatic_regime} considered here, this correction remains perturbatively small and stays subleading relative to the standard Hubble-friction term in the tensor mode equation, so we neglect it in what follows.

\paragraph*{Parity-even part (threshold corrections).}
In the heavy-mass regime $m(\phi)\gg H$, the dominant contributions to the inflationary EFT are local and can be captured by the same heat-kernel expansion~\cite{Achucarro:2012sm,Vassilevich:2003xt}.
\begin{equation}
\Gamma_{\rm even}=
\frac{1}{32\pi^2}\int d^{4}x_{E}\sqrt{g_{E}}\,\int_{0}^{\infty}ds
\sum_{n=0}^{\infty} s^{n-3}\,
\mathrm{tr}_V\!\big(a_{n}(x_E)\big)
\label{eq:even}
\end{equation} 
Unlike the parity-odd case, this proper-time integral is UV divergent at the lower limit. We regulate it by introducing a UV cutoff $\Lambda>0$,
\begin{equation}
\int_{0}^{\infty}ds\ \longrightarrow\ \int_{\frac{1}{\Lambda^2}}^{\infty}ds,
\end{equation}
which corresponds to a UV regulator in the effective action.
It is also convenient to factor out the dominant mass dependence by defining resummed coefficients $\tilde a_n$~\cite{Franchino-Vinas:2023wea,Barvinsky:1985an,Barvinsky:2002uf,Vassilevich:2003xt} through
\begin{equation}
e^{-s\,m^{2}(\phi)}
\sum_{n=0}^{\infty} s^{\,n-3}\,\mathrm{tr}_V\!\big(\tilde a_{n}\big)
\;\equiv\;
\sum_{n=0}^{\infty} s^{\,n-3}\,\mathrm{tr}_V\!\big(a_{n}\big).
\label{eq:resumm}
\end{equation}
In practice we use the left-hand side of Eq.~\eqref{eq:resumm} in Eq.~\eqref{eq:even} to render each term integrable at large $s$. 
This rearrangement is exact before truncation and therefore leaves the physics unchanged; truncating the series at finite 
$n$ is equivalent to expanding the EFT in a large-mass (derivative) series in powers of $1/m^{2}$.

In four spacetime dimensions, the proper-time integral scales as
\begin{equation}
\int_0^\infty ds\, s^{n-3} e^{-s m^2}\sim m^{4-2n},
\end{equation}
which immediately organizes the large-mass EFT expansion. Substituting the explicit \(a_n\) coefficients and performing the proper-time \(s\) integrals, one finds the following structure in the \(\Lambda^2\to \infty \) limit, up to an Euler--Mascheroni constant that can be absorbed into a redefinition of \(\Lambda^2 \).

The \(n=0\) term scales as \(m^4\) and contributes to the effective potential, and contains the expected quartic divergence \(\Lambda^4\), 
\begin{equation}
\Gamma_{\rm even}\!\supset\!
\frac{-1}{16\pi^2}\!\int\! d^4x\,\sqrt{-g}\,m^{4}\Bigg[\frac{\Lambda^4}{m^{4}}-\frac{2\Lambda^2}{m^{2}}-\ln\left(\frac{m^2}{\Lambda^2}\right)\!+\!\frac{3}{2}\Bigg].
\label{eq:even-0}
\end{equation}
The \(n=1\) term scales as \(m^2\) and renormalizes the Einstein--Hilbert term, and contains the quadratic divergence \(\Lambda^2\), 
\begin{equation}
\Gamma_{\rm even}\!\supset\!\frac{-1}{16\pi^2}\int d^4x\,\sqrt{-g}\,\frac{m^{2}R}{6}\Bigg[
\frac{\Lambda^2}{m^{2}}+\ln\left(\frac{m^2}{\Lambda^2}\right)-1\Bigg].
\label{eq:even-1}
\end{equation}
The \(n=2\) term is free of power divergences, and produces only logarithmic threshold corrections of the form \(\ln\!\big(m^2(\phi)/\Lambda^2\big)\), which renormalize dimension-four operators, including kinetic terms and higher-curvature invariants, 
\begin{equation}
\begin{split}
&\Gamma_{\rm even}\supset\!\frac{-1}{32\pi^2}\!\int\!d^4x\!\sqrt{-g}
\ln\!\left(\frac{m^2}{\Lambda^2}\right)\!
\Bigg[
2\partial_{\mu}m\partial^{\mu}m
\!+\!2m^{2}\partial_{\mu}\theta\partial^{\mu}\theta\\
&
\!+\!\frac{1}{72}R^{2}
\!-\!\frac{7}{360}R_{\mu\nu\rho\sigma}R^{\mu\nu\rho\sigma}
\!-\!\frac{1}{45}R_{\mu\nu}R^{\mu\nu}
\!-\!\frac{2}{3}q^{2}F_{\mu\nu}F^{\mu\nu}
\Bigg].
\end{split}
\label{eq:even-2}
\end{equation}
Terms with \(n\ge 3\) correspond to higher-derivative operators suppressed by extra powers of $1/m^2$, and are thus consistently neglected in the large-mass EFT expansion~\cite{Ellis:2020ivx,Barvinsky:1985an,Vassilevich:2003xt}.

Following the standard EFT procedure~\cite{Appelquist:1974tg,Henning:2016lyp,Ellis:2020ivx,Angelescu:2020yzf,Vassilevich:2003xt}, we remove all power divergences by the corresponding counterterms, and then fix the remaining finite part by a subtraction prescription at $\phi=\phi_\ast$, requiring the one-loop corrections vanish at the reference point, implemented by the replacement
\begin{equation}
\ln\!\left(\frac{m^2(\phi)}{\Lambda^2}\right)\;\to\;\ln\!\frac{m^2(\phi)}{m^2(\phi_\ast)}.
\end{equation}
Then all loop-induced corrections appear through the physical threshold logarithm \cite{Appelquist:1974tg,Henning:2016lyp}.
In particular, the effective potential and gauge kinetic term receive
\begin{equation}
\Gamma_{\rm even}\!\supset\!
    \frac{1}{16\pi^{2}}\!\ln\!\left(\!{\frac{m^{2}(\phi)}{m^{2}(\phi_*)}}\!\right)\!m^{4}(\phi)
+\frac{q^{2}}{48\pi^{2}}\!\ln\!\left(\!{\frac{m^{2}(\phi)}{m^{2}(\phi_*)}}\!\right)\! F_{\mu\nu} F^{\mu\nu}.
\label{eq:even-renormalize}
\end{equation}

The first term in Eq.~\eqref{eq:even-renormalize} is the Coleman--Weinberg threshold 
correction \cite{Coleman:1973jx} to the inflaton effective potential, whereas the second term encodes the one-loop vacuum polarization \cite{Schwinger:1948yj} of the $U(1)$ gauge field and hence a threshold renormalization of the gauge kinetic function \cite{Ellis:2020ivx,Schwinger:1948yj,Angelescu:2020yzf}. 
In standard axion case this correction is usually absorbed into a constant renormalization of the gauge coupling, but when the heavy-fermion mass depends on $\phi$ it becomes time dependent along the inflationary trajectory. In our setup,
\begin{equation}
\ln\!\frac{m^2(\phi)}{m^2(\phi_\ast)}
\simeq \frac{2\kappa}{M}\Big[T_{S}(\phi)-T_{S}(\phi_{*})\Big]+\mathcal{O}\Big(\frac{\kappa^{2}}{M^{2}},\,\frac{\sigma^{2}}{M^{2}}\Big).
\end{equation}

The remaining one-loop operators in Eqs.~\eqref{eq:even-0}-\eqref{eq:even-2} are subleading in the regime of interest. 
The derivative terms $2(\partial_\mu m)^{2}$ and $2m^{2}(\partial_\mu\theta)^{2}$ renormalize the inflaton kinetic sector. On the homogeneous background, they reduce to $2\dot{m}^{2}+2m^{2}\dot{\theta}^{2}$ and are thus suppressed relative to the canonical inflaton kinetic term by the squared adiabatic parameters $(\dot{m}/m^{2})^{2}$ and $(\dot{\theta}/m)^{2}$ in Eq.~\eqref{eq:heavy_adiabatic_regime}. The curvature-squared invariants (i.e. $R^2$, $R_{\mu\nu}R^{\mu\nu}$, $R_{\mu\nu\rho\sigma}R^{\mu\nu\rho\sigma}$) affect only the gravitational sector and are loop-suppressed by $H^{2}/M_{\mathrm{Pl}}^{2}$ relative to the Einstein--Hilbert term on the inflationary background.
The $m^{2}R$ operator likewise induces only a small field-dependent renormalization of the Einstein--Hilbert term, suppressed by $m^{2}/M_{Pl}^{2}$ in the small-modulation regime. Its effect on the inflaton dynamics is further suppressed relative to the Coleman--Weinberg term by $R/m^{2}\sim H^{2}/M^{2}$.
We therefore neglect these terms and retain only the leading corrections to $V(\phi)$ and $F_{\mu\nu}F^{\mu\nu}$.

Combining the tree-level action~\eqref{eq:tree} with the anomaly-induced parity-odd term~\eqref{eq:odd} and the retained parity-even threshold corrections~\eqref{eq:even-renormalize}, we arrive at the inflationary EFT in Eq.~\eqref{eq:L_axion}.

\section{Benchmark choice and relation to the UV completion}
\label{app:benchmark}

The benchmark EFT parameters appearing in Eqs.~\eqref{eq:I1}--\eqref{eq:Veff} of Sec.~\ref{sec: MODEL} can be related to the underlying heavy-fermion completion through
\begin{equation}
\delta=-\frac{q^{2}}{6\pi^{2}}\frac{\kappa}{M},
\qquad
\alpha=\frac{q^{2}}{4\pi^{2}}\frac{\sigma}{M},
\qquad
\gamma=-\frac{\kappa M^{3}}{8\pi^{2}}.
\end{equation}

In the numerical analysis, we choose the benchmark parameters in a way that is consistent with the regime of validity assumed in Appendix~\ref{app:oneloop}. We first fix
\begin{equation}
V_0=2.64\times10^{-11}M_{\mathrm{pl}}^{4},
\end{equation}
which sets a representative inflationary energy scale. We then take
\begin{equation}
M=3.74\times10^{3}\sqrt{V_0}\,M_{\mathrm{pl}}^{-1},
\end{equation}
so that the heavy fermion mass is hierarchically larger than the Hubble scale during inflation, ensuring that the heavy-field condition is well satisfied. Next, we choose
\begin{equation}
\frac{\kappa}{M}=-0.0153,
\qquad
\frac{\sigma}{M}=0.1,
\end{equation}
which lie in the small-modulation regime $|\kappa|,|\sigma|\ll M$ assumed in the one-loop expansion. Finally, we take
\begin{equation}
\beta=3.75\,M_{\mathrm{pl}}^{-1},
\qquad
q=58.94,
\end{equation}
guided by both phenomenological and consistency requirements. On the one hand, $\beta$ is chosen to be large enough to generate a sufficiently localized threshold feature for the phenomenology studied in the main text, while $q$ is taken large enough for the loop-induced corrections to the gauge-sector functions to have an appreciable effect. On the other hand, $\beta$ cannot be taken arbitrarily large, since the benchmark must remain compatible with the heavy, adiabatic regime in Eq.~\eqref{eq:heavy_adiabatic_regime}, in particular the slow-variation conditions
\(
|m'(\phi)\dot\phi|/m^2\ll1
\)
and
\(
|\theta'(\phi)\dot\phi|/m\ll1
\). Altogether, these values imply
\begin{equation}
\gamma=0.1V_0,\qquad
\delta=0.9,\qquad
\alpha=8.8,
\end{equation}
which are the EFT benchmark parameters quoted in Sec.~\ref{sec: MODEL}.

\bibliography{reference}


\end{document}